\documentclass[conference]{IEEEtran}
\usepackage{paralist}
\usepackage{lipsum}
\usepackage[cmex10]{amsmath}
\usepackage{makeidx}  
\usepackage{graphicx}
\usepackage{color}
\usepackage{url}
\usepackage[breaklinks]{hyperref}

\usepackage[capitalize]{cleveref}

\usepackage{algorithmic}

\graphicspath{{./figures/}}
\DeclareGraphicsExtensions{.pdf,.jpeg,.png,.PNG,.JPG}

\newcommand{\acknowledge}{This work was supported in part by
the Air Force Office of Scientific Research under the AFOSR
award FA9550-12-1-0476, by the National Science Foundation
grants NSF/OCI---0941434, 0904782, 1047772, and
by the U.S. Department of Energy, Office of
 Advanced Scientific Computing Research,  through the Ames Laboratory,
 operated by Iowa State University under contract No.~DE-AC02-07CH11358.}

\begin{document}

\title{Experimentation Procedure for Offloaded Mini-Apps \\ Executed on
Cluster Architectures with \\ Xeon Phi Accelerators}

\author{\IEEEauthorblockN{Gary Lawson$^{\dagger}$, Vaibhav Sundriyal,
Masha Sosonkina, and Yuzhong Shen}
\IEEEauthorblockA{Old Dominion University, Norfolk VA 23529, USA,\\
Department of Modeling, Simulation, and Visualization Engineering,\\
\{ glaws003, vsundriy, msosonki, yshen \} @odu.edu}}

\maketitle    

\begin{abstract}

A heterogeneous cluster architecture is complex. It contains hundreds, or thousands of devices
connected by a tiered communication system in order to solve a problem. As a heterogeneous system,
these devices will have varying performance capabilities. To better understand the interactions which
occur between the various devices during execution, an experimentation procedure has been devised
to capture, store, and analyze important and meaningful data. The procedure consists of various tools,
techniques, and methods for capturing relevant timing, power, and performance data for a typical
execution. This procedure currently applies to architectures with Intel Xeon processors and Intel Xeon
Phi accelerators. It has been applied to the Co-Design Molecular Dynamics mini-app, courtesy of the
ExMatEx team. This work aims to provide end-users with a strategy for investigating codes executed on
heterogeneous cluster architectures with Xeon Phi accelerators.

\end{abstract}

\section{Introduction} \label{sec:intro}
Measuring the performance and energy of an application can be a challenge. There are tools and
methods for obtaining power and performance measurements, but accurately combining these
with execution can be difficult. Further, from the point-of-view of a
single developer, determination of the critical and non-critical execution points
can be tedious or overwhelming. This work aims to provide an easy-to-reproduce
procedure for accurately profiling a generic application with minimal code changes.
This is beneficial to the solo developer looking to optimize an application, because
key phases of execution within the application will fall into the authors
outlined measurement scheme and promote isolation of these application phases.
From the point-of-view of co-design, this work provides meaningful insights
into the performance of an application; metrics such as memory bandwidth
and computational throughput are used in place of application phases to
describe execution time. However, obtaining these metrics does require execution,
hence a well described procedure has been developed.

Accelerators are often adopted to reduce time-to-solution with low
energy costs. From the work of Choi {\em et.~al}~\cite{Algorithmic}, the Xeon Phi
is capable of 11 GFLOPs/J and 880 MB/J for single-precision operations
(measured throughput of 2 Tflops/s and 180 GB/s memory bandwidth).
The Intel Xeon Phi is an accelerator that promotes high memory bandwidth
(i.e. data movement) in addition to high computational throughput, and that
supports various execution modes~\cite{XPoverview, micquickstart}. The Xeon Phi also offers user-level access to important
power data, but poorly document how to utilize the information; this work sheds light
on an easy-to-implement method to read power at the highest available sampling frequency
for the device. Further, unlike tools or methods which relay on reading ``window'' power,
such as MICSMC (Software Management Controller)~\cite{micsmc}, this
work yields true, instantaneous power measurements based on the connectors that supply power.

The Xeon Phi co-processor is an accelerator with many execution modes: {\em native}, {\em offload},
and {\em symmetric}. This is unique to accelerators because normal operation of an accelerator is
considered the {\em offload} execution mode; this is the mode used by GPU’s. The Xeon Phi supports
these other modes because of the micro-OS (running a special version of Linux) which enables
the device to execute applications from the device itself; to the host, the Xeon Phi may be
considered an additional node. Native execution mode allows an application to be executed only
on the Xeon Phi; a user log's onto the device and executes the application. Symmetric execution mode
allows an application to be executed on the host and on the Xeon Phi, but each device is to solve
a different sub-domain. Offload execution mode allows the host and accelerator to share the workload,
but this execution mode requires some code changes. This is the ideal starting place for a
procedure that requires some code instrumentation.

A heterogeneous cluster architecture is composed of many nodes connected by a network interconnect.
Each node is composed of multiple devices with varying performance capabilities. An application
executed on such an architecture must implement domain decomposition~\cite{domaindecomp}; to
take a problem and divide it into independent sub-problems, or sub-domains. These sub-domains are
then distributed to each node, one or more sub-domain per node. However, domain decomposition
requires data sharing between sub-domains to solve the total problem completely. For more simple
implementations, computation of the problem and communication between sub-domains do not
overlap. In this work, overlap between these two phases is not considered.

A heterogeneous cluster architecture is complex. Measuring the performance and energy consumption
of such an architecture is also complex. Ideally, a measurement procedure should have low
performance impact and energy overhead; measuring an execution should not degrade the performance of the
application, nor dramatically increase the required power draw. A procedure should also be easy-to-
implement on other systems, and should not require dramatic code changes. In this work, such a
procedure is presented to capture important data for all devices utilized in a heterogeneous cluster
architecture. Important data includes bandwidth for various data transfers, work performed by the Xeon
Phi accelerator, and power consumption for the various devices.

This work investigates specifically the offload execution mode~\cite{micoffload, micdatasheet},
however the procedure may be applied
to native and symmetric execution modes as well with minimal changes. Some of these changes are to
be introduced in this work, but have not been thoroughly tested; although this is a future work. To measure performance,
hardware counters are read using the Performance API (PAPI)~\cite{papi}. Host power is measured using
the Running Average Power Limit (RAPL) interface~\cite{rapl}, and Xeon Phi power is measured by
reading the power file “{\tt /sys/class/micras/power}”~\cite{micpower}.  Timings are
gathered during execution of an application; however additional timings are necessary beyond what is
provided by default. This work will explain the additional code changes required for an application to
provide accurate event timings to be used to synchronize execution flow with the power and
performance samples.

The remainder of this paper is organized as follows:~\cref{sec:prep} provides details for
cluster and application requirements, such as software, hardware, and code changes.
\Cref{sec:prof} discusses defining the experiment and post-execution data processing
involved in profiling execution output.   Finally, \cref{sec:conc} concludes.

\section{Initial Steps} \label{sec:prep}

Before experimentation may proceed, certain software and hardware are required to accurately
measure an execution. In general, the authors assume the cluster is composed of Intel~Xeon
processors of the Sandy-Bridge or newer micro-architecture supplied with one
or more Intel Xeon Phi accelerators.
This architecture configuration is assumed to be most common (among hybrid Xeon and Xeon Phi clusters),
and provides the necessary
hardware counters for model parameter estimations. However, it is not a requirement that
the processors be of the Intel brand or the accelerator be a Xeon Phi {\em if} similar
measurements may be obtained.

Beyond compiling the application~\cite{micintelcompiler} and executing in an MPI (Message Passing Interface)
environment~\cite{intelmpi}, the cluster architecture should have access to RAPL on the host, and
a version of PAPI for both the host and native Xeon Phi; even for offload execution.
Although the application will require calls to PAPI from within the offload section,
PAPI itself should be natively compiled and the library accessible during compilation
and execution of the application. For offload execution, a host version of PAPI is
also required because offload sections may be executed on the host in the event of
a conditional offload, or should ``no-offload'' be enabled during compilation. It is
also important to note that for native Xeon Phi hardware counters, only versions
5.3.0 and 5.3.2 currently support this functionality; more recent versions of PAPI
 (up to 5.4.1 as of this version) are unable to convert native hardware counter names
 into codes and therefore are unable to access the counter. Removing the
dependency of PAPI from the parameter estimation procedure will be done by the
authors in the future since PAPI provides limited support for the Xeon Phi.
It is important to note that with the more recent versions of MPSS, configuring
PAPI requires more than what is provided in the instructions for the PAPI
5.3.X versions. Specifically, the authors
followed the instructions from version 5.4.1~\cite{papi} to use the following
configurations options:
\begin{compactitem}
  \item {\tt --with-mic}
  \item {\tt --host=x86\_64-k1om-linux}
  \item {\tt --with-arch=k1om}
  \item {\tt --with-ffsll}
  \item {\tt --with-walltimer=cycle}
  \item {\tt --with-tls=\_\_thread}
  \item {\tt --with-virtualtimer= \\ \enspace \enspace clock\_thread\_cputime\_id} ,
\end{compactitem}
along with all the other configutation steps as per instructions to PAPI versions 5.3.X.

\subsection{Code Instrumentation} \label{subsec:codeinst}
This section defines the required code changes to implement offload execution in CoMD and to obtain
highly accurate timings for the computation and communication phases on each device. These timings
may then be cross-referenced to the power and performance measurements to define various
performance metrics, and total energy consumption. Associate code changes and micro-measurement
apps are provided. Each micro-measurement app only spawns a single-thread to sample and print.

\subsubsection{CoMD Overview}
CoMD is a proxy application
developed as part of the Department of Energy co-design research
effort~\cite{co-design} Extreme Materials at Extreme Scale~\cite{exmatex}
(ExMatEx) center. CoMD is a compute-intensive application where approximately 85-90\%
of the execution time is spent computing forces.
Although two methods are available for the force computation, this work
focuses only on one of them, the more complex and accurate EAM force kernel
for short-range material response simulations, such as uncharged metallic
materials~\cite{comd}. The EAM kernel was chosen because its parallel
performance generally receives less attention than the more commonly used
Lennard-Jones potential, which easily yields itself to parallelism.

\subsubsection{Setup of the Offload Execution Mode} \label{subsubsec:offdec}
Offloading to MIC requires the use of special pragmas defined for the Intel compilers. These
pragmas specify the code sections to be processed by the Xeon Phi accelerator. Within the pragma
statement, one must specify the MIC device to communicate with, the data to be transferred with the
associated parameters (such as array length, data persistence, variable reassignment, etc.), offload
conditional, and whether the offload event is asynchronous, among other
options~\cite{micintelcompiler}. However, in addition to simply specifying what code sections to process
on the Xeon Phi, the arrays must be properly formatted for optimal transfer bandwidth.

It is possible, although inefficient, to transfer multi-dimensional arrays between the host and Xeon Phi.
Therefore algorithm structures should conform to the structure-of-arrays data layout; CoMD is originally
organized as an array-of-structures which do not transfer easily. This code change simply requires
converting the multi-dimensional arrays into one-dimensional arrays. In the most recent experiments,
CoMD is measured to obtain more than 3 GB/s bandwidth over the PCI bus (of 8 GB/s) for a problem
size of 50 (500,000 atoms); the resulting communication time is insignificant with respect to the
remaining computational requirements. This is one of many measurements to be obtained using the
procedure. One final code improvement is the re-assignment of the maximum number of atoms per link
cell: by default, a link cell may contain 64 atoms but has been reduced to 16 to reduce memory
requirements per thread. PAPI must be instrumented into the offload sections such that
memory and bandwidth may be approximated during execution. PAPI is simply started and stopped
for each offload section such that the counter is always reset for the next offload; the result
is printed with application output. SSE3 instructions have been enforced during
compilation~\cite{sseinst}, and utilization of the 2 MB buffers available through
the environment variable~\cite{mic2mbbuffer} has been implemented.

\subsubsection{Synchronization of Measuring Event Timers} \label{subsubsec:datasynch}
Power and performance measurements are obtained for each device individually. This approach
removes unnecessary overhead because devices are not required to communicate measurement data
during execution. However, this approach also requires the use of three (or more) separate timers: the
algorithm timer, host timer, and accelerator timer(s). To synchronize these data files, two timings are
output with each measurement or event output statement: local time in the format [HH:MM:SS], and
the time from start as a decimal value. The time from start (TFS) value represents the time elapsed from
the start of each measurement tool or algorithm execution. The use of the local timestamp ensures all
timings are accurate to within one second, however in addition to TFS, the error in timings is reduced to
a fraction of a second (within 20ms for the host, and 100ms for the Xeon Phi).

Until a more sophisticated, and automatic method is developed, direct source code manipulation is the
simplest solution to start with. However, as offload execution already requires source code
manipulation, the additional event timing statements are reasonable; especially for CoMD which
features robust profiling output by default. Specifically, CoMD provides four specific functions which
must occur within one iteration of the simulation; for other codes one or more functions may be
required, but in general it is most important to quantify total simulation time, the time spent performing
offload execution, and the time spent transferring data during the communication phase.

\subsubsection{Obtaining Execution Time Values}
The execution timings of interest are specifically: the time to compute on the host,
communicate on the host, compute on the Xeon Phi, and transfer data over the PCI bus.
To obtain these timings, the application output must be consulted: for CoMD, the
timings are excellently profiled although only the root timings are provided in
entirety. For other applications, additional timings may be required to obtain
host computation; this will be investigated in the future. The execution timings,
which are already collected for each sub-domain, have been exposed to obtain
exact timings for each sub-domain. Although this is not so much of interest in
these offload-only execution experiments; the authors preliminary investigation
into executions with multiple execution modes showed this information to be
crucial and thus has been maintained for future investigations until an
improved method has been determined.

\subsubsection{Using CPU Hardware Counters}
The host CPU micro-measurement application has been developed to continuously read the
RAPL power counter for CPU Core and DRAM power; the sum is regarded as total CPU power
as uncore device power is not considered. Additionally, host performance is measured
using PAPI where the last-level-cache memory fill counter, and unhalted CPU cycles are
measured; the hardware counter name differs slightly depending on micro-architecture.
For Sandy-Bridge, ``{\tt MEM\_LOAD\_UOPS\_MISC\_RETIRED:LLC\_MISS}'', and
``{\tt CPU\_CLK\_UNHALTED:THREAD\_P}'' are the native hardware counters used to approximate
host memory usage and bandwidth. For Ivy-Bridge, unhalted cycles are captured using
the same counter as on Sandy-Bridge, but the LLC memory counter is:
``{\tt MEM\_LOAD\_UOPS\_RETIRED:L3\_MISS}''. Power and performance are sampled at a rate of
10 m$s$; the resulting data is printed to output with the timestamp and TFS for
synchronization.

\subsubsection{Using MIC Hardware Counters}
The MIC\footnote{{\em MIC} stands for ``many-integrate cores'' technology used
in Intel Xeon Phi.} micro-measurement application has been developed to continuously read
the available power file~\cite{micpower}: {\tt /sys/class/micras/power} which provides approximated
power over two time windows, power to each connector, and voltage and power readings
to the core, uncore, and DRAM devices. Unlike the host CPU definition of power,
MIC power is based on the power draw measured for each connector: PCI-E, 2x3,
and 2x4. The sum of each connector is the absolute power draw for the device
as defined in the Xeon Phi data sheet~\cite{micdatasheet}. The power file is updated
only every 50ms, thus is the lowest available sample rate for the device.

For offload execution,
the MIC micro-measurement app only measures power; however for native or symmetric
execution, this micro-app would also measure performance with PAPI. For the Xeon Phi,
the native hardware counters of interest are: ``{\tt L2\_DATA\_READ\_MISS\_MEM\_FILL}'' and
``{\tt CPU\_CLK\_UNHALTED}''; these are used to obtain estimates for memory usage and
bandwidth. For offload execution, these hardware counters are instead measured over
the duration of each offload section and print with application output.
To obtain an estimate for vectorization intensity,
a few executions (although one is really all that is needed) using the
hardware counters: ``{\tt VPU\_ELEMENTS\_ACTIVE}'' and ``{\tt VPU\_INSTRUCTIONS\_EXECUTED}''
should be measured, where elements over instructions equates to vectorization intensity.
In general, the value should be between 1 and 8 for double-precision, and
1 and 16 for single-precision~\cite{optperf2}.

\section{Profiling Executions} \label{sec:prof}
In this work, an execution is more than simply running the application; it
requires properly measuring CPU and Xeon Phi power and performance, and
synchronizing this output with the applications execution. A properly
defined experiment must be presented, and the process of mining the raw
output data is also discussed. The result of this process are measured
performance metrics to describe various attributes for an application,
such as the total workload for the accelerator defined in FLOPS and
bandwidths for many different data transfer situations. The executions
are always run with the {\tt offload report} environment variable set to 2;
this provides MIC time, CPU time (if available), and data transferred to
and from the device. To distinguish offload reports between various sub-domains,
it is advised that MPI is executed with the `{\tt -l}' option to print the sub-domain
identification number~\cite{mpioffloadreports}.

\subsection{The Experiment}
The experiment should be designed such that all investigated parameters are meaningful.
In this work, six meaningful parameters have been chosen: the system, number of nodes,
number of Xeon Phi per node, total problem size, host frequency (no DVFS), and
Xeon Phi cores used. In general, the authors are interested in determining the
optimal configuration (defined by all six parameters) which is defined by a
{\em static configuration} set (defined by: system, nodes, MICs/node, and problem size)
and {\em configuration space} (defined by: host frequency and MIC cores).
On the Borges system, two static configurations per problem size are investigated:
{\em MIC 1} and {\em MIC 2}, because the system consists of a single-node. On Bolt,
six static configurations per problem size are investigated: {\em N\# MIC \#}; Bolt
only has three nodes with two Xeon Phi, hence one, two, and three node configurations are
investigated, each with one and two Xeon Phi used. The parameters have been grouped
into static configurations and configuration space because static configurations may be
easily compared with one another and defined with a minimum energy; the minimum energy
may be found in the configuration space, because these parameters impact execution
energy and performance. Note, although the number of Xeon Phi also impacts energy,
it is often desirable to compare the performance and energy for each investigated.

An experiment is composed of many executions; each may vary in configuration, but each follows the
same execution process to ensure minimal measurement overhead. The process for a typical execution
is as follows:
\begin{compactenum}
  \item Start CPU micro-measurement app on all nodes
  \item Start MIC micro-measurement app on all Xeon Phi
  \item Sleep 20 seconds
  \item Execute CoMD
  \item Sleep 10 seconds
  \item Stop MIC micro-measurement app on all Xeon Phi
  \item Stop CPU micro-measurement app on all nodes
  \item Copy MIC power output files from MIC to storage
  \item Sleep 60 seconds.
\end{compactenum}

Ample idle time is provided before execution begins to ensure a sufficient number of power samples
may be obtained for each device such that idle power may be measured. For larger clusters, the timing
for step three may need to be adjusted. The command to start each MIC power measurement is issued
using SSH which incurs a slight delay before power measurements may begin. Idle power measurements
are based on at least 10 seconds of sample data.
CoMD is then executed according to the execution configuration parameters. Upon completion, a brief
idle period is provided to capture power measurements before the CPU and MIC measurement threads
are halted. Finally, a rest period of one minute is provided to allow the system to cool-down. Ideally this
should be longer, but a typical experiment consists of hundreds or thousands of executions. The time
spent allowing the system to rest accounts for the majority of the total execution time for an
experiment.

\subsection{Post-Execution Profiling}
Upon successful completion of the experiment, a plentiful number of output files are available for
post-execution processing. This process involves synchronizing measurement output with
application execution to properly quantify timings, power measurements, and performance
for various phases and states involved in execution. These raw metrics are then to be used to
establish estimated global parameters that define each static configuration.

\subsubsection{Obtaining Execution Time Values}
Extracting execution time is fairly simple with CoMD. There are four main functions
that occur every iteration: update position, velocity, compute force, and share
data between sub-domains; for EAM, there is an additional data transfer during
the force computation since it is a more specific algorithm. Host execution time
is defined as time to update position plus velocity plus data redistribution
minus the data transfer occuring within the redistribution phase. Host communication
time is the sum of both communication phases (within the redistribution and force
phase), but is more accurately defined as two comm timings: halo exchange and
reduce. This is preferred because it is more interesting with respect to optimization
to separate the point-to-point data transfers and reduce function timings.
Xeon Phi computation time is based on the sum of offload report MIC time for
all offloads throughout the simulation. PCI transfer time is based on
the offload report as well; however is the difference between the sum of
CPU time and MIC time over all offloads during the simulation. If CPU time
is undefined (reports 0.0000 seconds), PCI time may be approximated as the
difference between total simulation time (defined as loop in the CoMD output
timings), and the time to compute and communicate on host and to compute on
the accelerator.

\subsubsection{Obtaining Power Consumption Values}
Extract power draw for each state, idle or active, is accomplished slightly
differently for each device because idle time and duration differs for each
device. For the host, the active state is defined by the host computation
time; for host communication, PCI data transfer, or computation on the
accelerator, the host remains idle. For the Xeon Phi, the active state is
defined by the time to compute on the accelerator, and the device is
otherwise idle. To synchronize power draw to execution state, the
TFS for each file has to be synchronized to the TFS value from the application
for the key execution phases. Host idle power is accumulated during host
communication and the entire force computation because it includes
accelerator computation and PCI data transfer. All other power is
accumulated in host active power. For the Xeon Phi, active power is
accumulated during offload execution and otherwise idle power is
accumulated with the current power sample. Power is finally divided by the
number of samples for each state (always greater than 100 samples).

\subsubsection{Obtaining Performance Values}
Extracting the performance metrics for each phase of execution is fairly
straight forward with the required code instrumentations. For the
Xeon Phi, memory usage is simply the LLC\_MISS counter multiplied by 64
bytes per cache line; bandwidth is memory usage times frequency divided
by unhalted clock cycles. Because these are measured explicitly for the
offload sections, these may simply be summated over all offloads
during the simulation. For the host, performance samples must first
be synchronized with execution and within the appropriate phase, but
because only the host communication phase is of interest with respect
to performance, this is the only phase in which the counters are accumulated.
The host follows the same simple formula for computing memory usage
and bandwidth as on the Xeon Phi. PCI memory usage is summated
over each offload report within execution of the simulation by
adding together the data sent to and recieved from the device.
PCI bandwidth is estimated by the amount of data to transfer
divided by transfer time.

Finally, work on the Xeon Phi may be
estimated by multiplying computational throughput and computation time.
Computational throughput may be calculated as the product of the
number of cores, vectorization intensity, average number of operations
per cycle, and operational frequency. For CoMD, vectorization
intensity is measured to be 2.6; operations per cycle is estimated
to be 1.15 as few fused-multiply operations are vectorized for this
version of CoMD. These two values depend heavily on the implementation
and application and must be measured and approximated approiately for
each application. Operations per cycle may be approximated by cross-referencing
the compiler vectorization report and source code to determine which
fused-multiply operations have been vectorized; all other operations
count as one. Then, assuming each operation were to count only as one,
take the ratio of number of vectorized to non-vectorized operations.

\section{Conclusions and Future Work} \label{sec:conc}

This work has provided a detailed procedure through which developers may
profile applications offloaded to accelerators and produce meaningful
conclusions and insights. At this stage, only the mini-application CoMD
has been investigated thoroughly with the procedure, but it is of the
utmost importance to validate the technique on many other mini-applications in
the future.
Additionally, reducing the number of executions to accurately measure
the application is a high priority because this would provide larger datasets
in a fraction of the time. Currently, the experiment requires several days
to complete in the cluster environment because each configuration change
requires a new execution and the associated system cool-down time.
Finally, collecting hardware counter measurements without the aid of
PAPI is also to be investigated.

\section*{Acknowledgments}

\acknowledge

\bibliographystyle{plain}
\bibliography{expproc15}

\end{document}